\documentclass[aps,preprint,showpacs,superscriptaddress]{revtex4}
\usepackage{amssymb}
\usepackage{bm}
\usepackage{graphicx}
\def \b{\begin{equation}}
\def \e{\end{equation}}
\def \ba{\begin{array}}
\def \ea{\end{array}}
\def \be{\begin{eqnarray}}
\def \ee{\end{eqnarray}}

\begin{document}
\title{Baryon deceleration by strong chromofields in ultrarelativistic nuclear collisions}%
\author{Igor N. Mishustin}
\affiliation{Frankfurt Institute for Advanced Studies, J.W. Goethe
University, D--60438 Frankfurt am Main, Germany} 
\affiliation{The Kurchatov Institute, Russian Research Center, 123182 Moscow, Russia}
\author{Konstantin A. Lyakhov}
\affiliation{Frankfurt
International Graduate School for Science, J.W. Goethe
University, D--60438 Frankfurt am Main, Germany}

\begin{abstract}
It is assumed that strong chromofields are generated at early stages of 
ultrarelativistic heavy-ion collisions which give rise to a collective 
deceleration of net baryons from colliding nuclei. We have solved classical
equations of motion for baryonic slabs under the action of a
time-dependent longitudinal chromoelectric field. It is demonstrated that the slab
final rapidities are rather sensitive to the strength and decay time of 
the chromofield as well as to the back reaction of the 
produced partonic plasma. The net-baryon rapidity loss $\langle\delta 
y\rangle=2.0$, found for most central Au-Au collisions at RHIC, can be
explained by the action of chromofields with the initial energy density 
of about 50 GeV/fm$^3$. Predictions for the baryon stopping at the LHC are 
made.
\end{abstract}

\pacs{25.75.-q, 12.38.Mh, 24.10.Jv, 24.85.+p}

\maketitle
It is expected that strong chromofields can develop at
early stages of ultrarelativistic heavy-ion collisions. There exist
different suggestions concerning the space-time structure of these
fields, ranging from string-like configurations as in the color Flux Tube
Model (FTM) \cite{Cash} to stochastic configurations associated with
the Color Glass Condensate (CGC) \cite{McL}. This picture is most conveniently
presented in the c. m. frame where two Lorentz contracted nuclei
look as thin sheets. After their intersection these sheets acquire
stochastic color charges as a result of multiple soft gluon
exchange. Then strong chromofields are generated in the space
between the receding sheets. At later times these fields decay into quarks
and gluons which after equilibration form a quark-gluon plasma. This
process has been studied by several authors under different
assumptions about the field decay mechanism (see e. g. Refs.
\cite{Gat,Kov,Gel}).

Most previous calculations assume that after interaction the nuclear debris
follow the light-cone trajectories, thus disregarding their energy
losses to produce the chromofield. This assumption can be possibly
justified only at asymptotically high energies. Furthermore, this
assumption becomes irrelevant when studying the baryon stopping.
Obviously, the energy of produced fields and particles is taken
entirely from the kinetic energy of the colliding nuclei.

As measured by the BRAHMS collaboration \cite{BRAHMS}, in central
Au+Au collisions at highest RHIC energy $\sqrt{s_{NN}}=200$ GeV 
per $NN$-pair the baryon energy losses are very significant, about 
70\% of the initial energy. The net-baryon rapidity distributions are
substantially shifted toward the center of mass from the initial
rapidities $\pm y_0$.

The problem of baryon stopping at RHIC has been addressed recently
by several authors. In particular, net-baryon rapidity distributions
were calculated within the microscopic string-based models like
UrQMD \cite{Bas1} and QGSM \cite{Ame}. Although these models
implement energy and momentum conservation, and thus predict a
certain baryon stopping, they are formulated in momentum space and
do not give a space-time picture of this process. Also, they are
dealing with hadronic secondaries and therefore preclude the
quark-gluon plasma formation. Recently, the calculations have 
been also done within a parton cascade model \cite{Bas2}.
The distribution of valence quarks at
central rapidities was also studied in ref. \cite{Ita} within a QCD
motivated approach which however can not be extended to the
fragmentation regions. 

In ref.\cite{Mis} a simple space-time model
was proposed where the baryon stopping was directly linked to the
formation of strong chromofields. There nuclear trajectories were
calculated under assumption that the field is neutralized at a sharp
proper time 1 fm/c (see also refs. \cite{Mag,Fri}). In this paper 
we further develop this model and apply it for a more realistic decay 
pattern of the chromofield. Our main goal is to calculate the net-baryon
rapidity distribution and compare it with the BRAHMS data.


Following ref. \cite{Iva} we decompose the collision of two
ultrarelativistic nuclei into a set of pairwise collisions of
elementary slabs. At given impact parameter ${\vec b}$ the positions of the 
target and projectile slabs in the transverse plane are determined by vectors 
${\vec s}$ and ${\vec b}-{\vec s}$, respectively. The transverse cross section area 
of individual slabs is assumed to coincide with the nucleon-nucleon inelastic 
cross section $\sigma_{NN}$. The energy and momentum of the projectile ($a=p$) 
and target($a=t$) slabs are parameterized as $E_a=M_a\cosh{Y_a}$ and 
$P_{a}=M_{a}\sinh{Y_{a}}~$, where $M_a=m_{\perp}N_a$ is the average transverse mass 
and $N_a$ is the average baryon number of slab $a$.
The average baryon transverse mass $m_{\perp}$ differs from the free nucleon mass 
$m_N$ due to internal excitation of slabs acquired at the interpenetration stage. 
It is parameterized in terms of the mean transverse momentum as 
$m_{\perp}=\sqrt{m_N^2+\langle p_{\perp}\rangle^2}$. The calculations below are made
in light-cone coordinates, proper time $\tau=\sqrt{t^2-z^2}$ and space-time rapidity 
$\eta=\frac{1}{2}\ln{\left(\frac{t+z}{t-z}\right)}$.   

In the Glauber model (see e.g. refs. \cite{Wong,Nardi}) the average number of 
participating nucleons from nucleus $a$ in a slab of transverse area $\sigma_{NN}$ 
located at a radius-vector ${\vec s}$ is given by the expression
\be \label{np}
N_a({\vec b},{\vec s})=\sigma_{NN} A_a T_a({\vec b}-{\vec s})\left(1-\left[1-\sigma_{NN}
T_{\overline{a}}({\vec s})\right]^{A_{\overline{a}}}\right)~,
\ee 
where $\overline{a}=t$ for $a=p$ and vice versa, $A_a$ is the mass number of nucleus $a$.
The normalized functions $T_a$ ($a=p,t$) describe the 
transverse profiles of the baryon number distribution in colliding nuclei. 
They are obtained by the integration of the corresponding baryon densities along 
the beam direction, $A_aT_a=\int\rho_a({\vec r})dz$.
The average number of nucleon-nucleon collisions in an inelastic 
interaction of two slabs at a radius-vector ${\vec s}$ is expressed as
\be \label{ncol}
N_{\rm coll}({\vec b},{\vec s})=\sigma_{NN}^2A_p T_{p}({\vec b}-{\vec s})
A_t T_{t}({\vec s})~.
\ee

We assume that the initial energy density
stored in the chromofield after collision of two baryonic slabs can
be parameterized as 
\be 
\label{field} 
\epsilon_f({\tau_0;\vec b},{\vec s})=\epsilon_0\left(\frac{s}{s_0}\right)^\alpha\left [N_{coll}(\vec{b},\vec{s})\right ]^{\beta}~,
\ee 
where $\epsilon_0$ is the fitting parameter, second factor describes the collision energy 
dependence with $\alpha\approx 0.3$, as motivated by small x behavior of the gluon structure
function \cite{Lev}. The last factor describes effects of the collision geometry, where exponent $\beta$ 
is related to the spatial distribution of the chromofield. For non-overlapping strings $\beta\approx 1.0$, but in the case of string clustering $\beta$ is expected to be closer to $0.5$ \cite{Bra}. 
In this paper we consider only the case $\beta=1$.

The equations of motion for a baryonic slab under the action of the longitudinal 
uniform chromofield are obtained by applying the energy-momentum conservation 
laws across the slab. It is assumed that the space between the receding slabs 
is occupied by the chromofield and a partonic plasma, produced by the partial field 
decay. They exert a certain force on the slab from inside. But from the other 
side the slab has nothing but the physical vacuum. This results in a net
force acting on the slab.  
The final system of differential equations governing the slab rapidities
$Y_a(\tau)$ has the following form (detailed calculations see in ref. 
\cite{Mis2})
\be
\frac{d{\tilde P_a}}{d\tau}=\mp B(\tau)-\frac{{\tilde P_a}}{\tau}, \label{P}\\
\frac{dM^2_a}{d\tau}=\mp 2A(\tau){\tilde P_a},\\
\frac{d\eta_a}{d\tau}=\mp\frac{{\tilde P_a}}{\tau{\tilde E_a}}, 
\ee
where $\tilde{P}_a=M_a\sinh{(Y_a-\eta_a)}$, ${\tilde E_a}=\sqrt{M_a^2+{\tilde P_a}^2}$,
$B(\tau)=\sigma_{NN}\left(\epsilon_{vac}+\epsilon_f-p\right)$ and $A(\tau)=\sigma_{NN}\left(\epsilon_p+p\right)$. 
The plus and minus signs in the r. h. s. of these equations correspond to the 
projectile ($a=p$) and target ($a=t$) slabs, respectively. We adopt the initial
conditions for the slab trajectories $\eta_{a}(\tau_0)=Y_{a}(\tau_0)=\pm y_0$
at $\tau_0$=0.01 fm, where $\pm y_0$ are the initial c.m. rapidities of colliding 
nuclei. In expressions above 
$\epsilon_p$ and $p=c_s^2\epsilon_p$ are, respectively, the energy density 
and pressure of the partonic plasma, and $c_s$ is the sound velocity.
The vacuum energy density $\epsilon_{vac}$ (bag constant) is introduced to 
account for the fact that the chromofield and partonic plasma can exist only in 
the perturbative vacuum. In numerical calculations $\epsilon_{vac}$ is fixed to 
the value 0.4 GeV/fm$^3$. It is assumed that the chromofield energy density 
$\epsilon_f(\tau)$ and the plasma energy density $\epsilon_p(\tau)$ are functions 
of the proper time $\tau$ only, defined in the interval 
$\eta_t(\tau)\leq\eta\leq\eta_p(\tau)$. The plasma energy density, $\epsilon_p(\tau)$ 
is found from the hydrodynamical equation with a source term due to the field decay:  
\be \label{hydro}
\frac{d\epsilon_p}{d\tau}+\left(1+c_s^2\right)\frac{\epsilon_p}{\tau}=
-\frac{d\epsilon_f}{d\tau}~.
\ee

We integrate the Eqs. (4)-(6) until the
time when the total pressure in the region between the slabs
vanishes, i.e. $B(\tau)=0$. After this time the slabs move with
constant velocity. Obviously, the solution of equation (\ref{P}) is
\be
{\tilde P_a}=\tilde
P_a(\tau_0)\frac{\tau_0}{\tau}\mp\frac1{\tau}\int\limits_{\tau_0}^{\tau}B(\tau)\tau
d\tau~,
\ee
and the slab rapidity $Y_a(\tau)$ can be easily found from the
definition ${\tilde P_a}=M_a\sinh{(Y_a-\eta_a)}$. Then, the slab trajectory, $z_a(\tau)$,
is obtained from the relation $z_a=\tau\sinh{\eta_a}$.


Below we present results for Au+Au and d+Au collisions at maximum RHIC 
energy $\sqrt{s_{NN}}$=200 GeV per NN pair ($y_0=5.4$).
The nucleon-nucleon inelastic cross section was taken to be
$\sigma_{NN}=42$ mb \cite{Nardi}. 
The baryon mean transverse momentum $\langle p_{\perp}\rangle$,
which determines the internal energy of baryonic slabs,
was fixed at the value 1.0 GeV/c, in accordance with BRAHMS data 
\cite{BRAHMS} for midrapidity region.

We have considered several functional forms for the time dependence of the
chromofield, resulting in different plasma production rates and baryon deceleration 
patterns. Here we present results for the power law, $\epsilon_f(\tau)=
\epsilon_f(\tau_0)\left[1+\frac{\tau-\tau_0}{\tau_{d}}\right]^{-4}$ characterized by 
the decay time $\tau_{d}=0.6$ fm/c, as expected for the Schwinger-like decay mechanism 
\cite{Gat}). 
This function is shown in Fig.~1 together with the time dependence of the plasma energy 
density as predicted by the hydrodynamical equation (\ref{hydro}). 
According to our picture, the Quark-Gluon Plasma(QGP) is produced by the continuous 
transformation of the field energy into the quark-antiquark and gluon pairs. Because 
of the delayed production, the plasma energy density is always smaller than the 
initial energy density of the chromofield. In accordance with previous calculations \cite{Gat},
we find that it reaches only about $22\%$ of the latter for the Schwinger-like decay law.

\begin{figure}
\includegraphics[width=14cm]{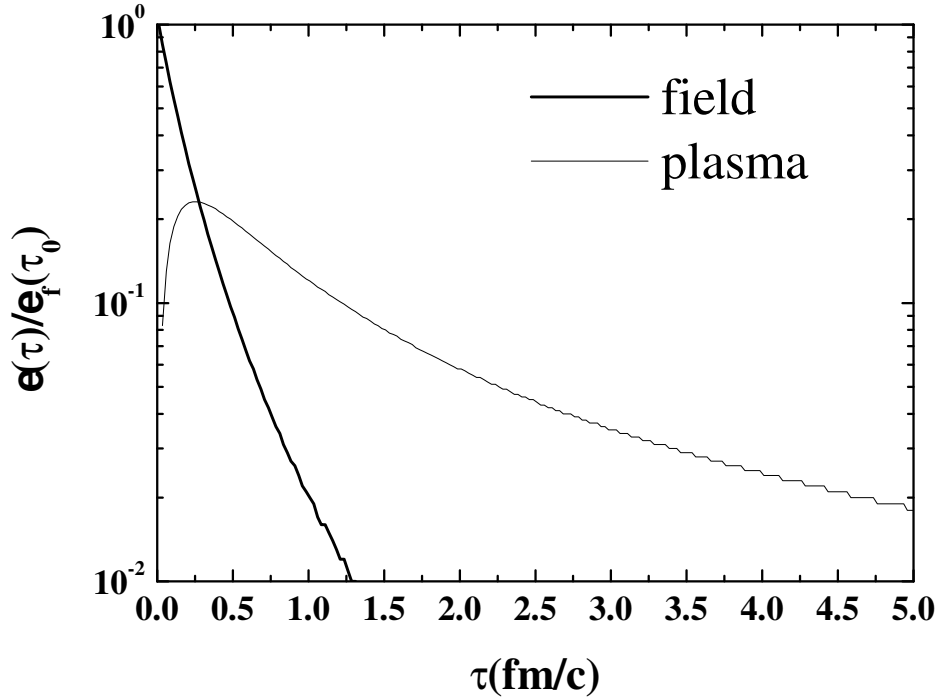} 
\vspace*{-0.5cm}
\caption{Evolution of the chromofield energy density (thick solid line) and QGP energy density
(thin solid line) in units of the initial chromofield energy density 
$\epsilon_f(\tau_0)$. Results are shown for the Schwinger-like (power-law) 
decay mechanism with  $\tau_d=0.6$ fm/c.}
\end{figure}

Figs.~2 shows how the time evolution of the slab rapidities $Y_{p,t}(\tau)$ depends 
on the initial energy density of the chromofield, as well as on the baryon numbers of 
colliding slabs. The ideal equation of state ($c_s^2$=1/3) was used for the 
quark-gluon plasma. We present results for several values of the parameter 
$\epsilon_0$ (see eq. (\ref{field})) between 0.5 and 2.0 GeV/fm$^3$. 
Upper panels show results for the collision of two equal slabs with 
baryon numbers $N_p=N_t$=5.8, representing an average pair of slabs in a central 
Au+Au collision.  The initial energy densities of the chromofield, $\epsilon_f(\tau_0)$,
range in this case from 16.8 (lower curves) to 67.3 (upper curves) GeV/fm$^3$. 
According to the BRAHMS data \cite{BRAHMS} the mean baryon rapidity
loss for the most central Au-Au collisions is $\langle \delta y\rangle\approx 2.0$. 
From the figures one can see that slab rapidities $Y_{p,t}(\tau)$ drop rapidly
at very early stages of the deceleration process, when the field is still strong. 
At later times, not only the decreasing chromofield (see Fig.~1) but also the plasma 
counter-pressure (left panels) cause an early termination of the deceleration 
process. As the result, the asymptotic slab rapidities have a low sensitivity to 
the initial chromofield energy density. Nevertheless, the observed rapidity loss 
can be achieved with the initial field energy density $\epsilon_f(\tau_0)$=50$\div$
70 GeV/fm$^3$. On the other hand, it is likely that the plasma pressure in the vicinity 
of the slabs is not as strong as in the central region. Then the deceleration process 
will be less affected by the plasma, and resulting slabs' rapidities will be smaller.
This is illustrated in the right panels showing calculations of the slab rapidities ignoring 
the plasma pressure. In this case the required rapidity loss can be achieved with a 
much weaker chromofield, $\epsilon_f(\tau_0)\approx$ $30\div 40$ GeV/fm$^3$. 
It is clear that the realistic situation will be in between of these two extremes. 

\begin{figure}
\includegraphics[width=15cm,height=15cm]{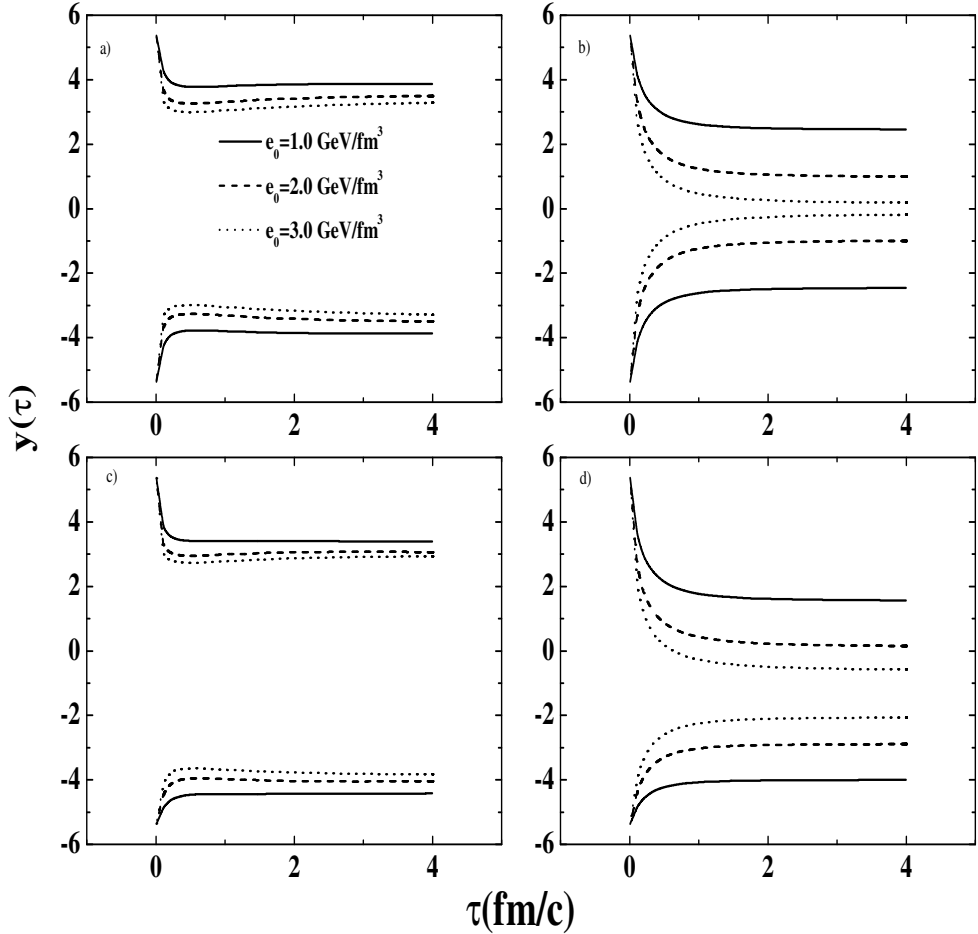} 
\vspace*{-0.5cm}
\caption{Projectile (upper curves) and target (lower curves) slab rapidities as functions 
of proper time calculated for the power-law chromofield decay with $\tau_d$=0.6 fm/c. 
Different pairs of curves correspond to different values of the parameter $\epsilon_0$ 
(indicated in the figure).
Results are shown for two cases: a,b) equal slabs with $N_p=N_t=5.8$ 
representing a central Au+Au collision, and c,d) two different slabs with 
$N_p$=2, $N_t$=8.8 representing a central d+Au collision. Left and right panels show 
the calculations with and without the back reaction of the produced plasma, 
respectively.}
\end{figure}

Low panels show results for an asymmetric collision of $N_p$=2
and $N_t$=8.8 slabs, which can be interpreted as a deutron colliding with a center
of a gold nucleus. The parameter $\epsilon_0$ is the same as before but now it 
corresponds to $\epsilon_f(\tau_0)$ between 8.9 and 35.5 GeV/fm$^3$. 
One can see that the rapidity shifts are now different for the 
light and heavy slabs. Moreover, the rapidity lost by the smaller slab is significantly  
larger as compared with the bigger one. This is of course a direct consequence of the fact 
that equal forces generate a larger deceleration for a smaller body. It is interesting 
to note that the mean rapidity loss of a deutron-like nucleus in a d+Au collision is 
predicted to be about two times larger than the mean rapidity loss of the gold nuclei
in a central Au+Au collision. 

In Fig.~3 we present the net-baryon rapidity distributions calculated for the 
central Au+Au collisions at maximum RHIC energy $\sqrt{s_{NN}}$=200 GeV per $NN$-pair.
The calculations were done by applying Gaussian weights for different color charges 
generated on the slabs \cite{McL}. Since the energy density of 
the chromofield is proportional to the square of the areal charge density, 
like in a capacitor, these weights are exponential in terms of the field energy 
density, i.e. $P(\epsilon_f)\propto \exp{\left[-\frac{\epsilon_f}{\langle\epsilon_{f}\rangle}\right]}$, where $\langle\epsilon_f\rangle\equiv
\epsilon_f(\tau_0)$ is the mean energy density of the field
 as parameterized in eq. (\ref{field}). The introduction of the fluctuating field 
leads to a very interesting effect: although the zero field has a maximum probability,
quite large values of the field, comparable with $\langle\epsilon_f\rangle$, 
are possible with a certain probability too. The larger fields lead to larger 
decelerations of the slabs. As a result, the $dN_B/dy$ distribution has a long 
tail towards central rapidities. This is exactly what is observed by the BRAHNS 
collaboration \cite{BRAHMS}. 
\begin{figure}
\includegraphics[width=14cm]{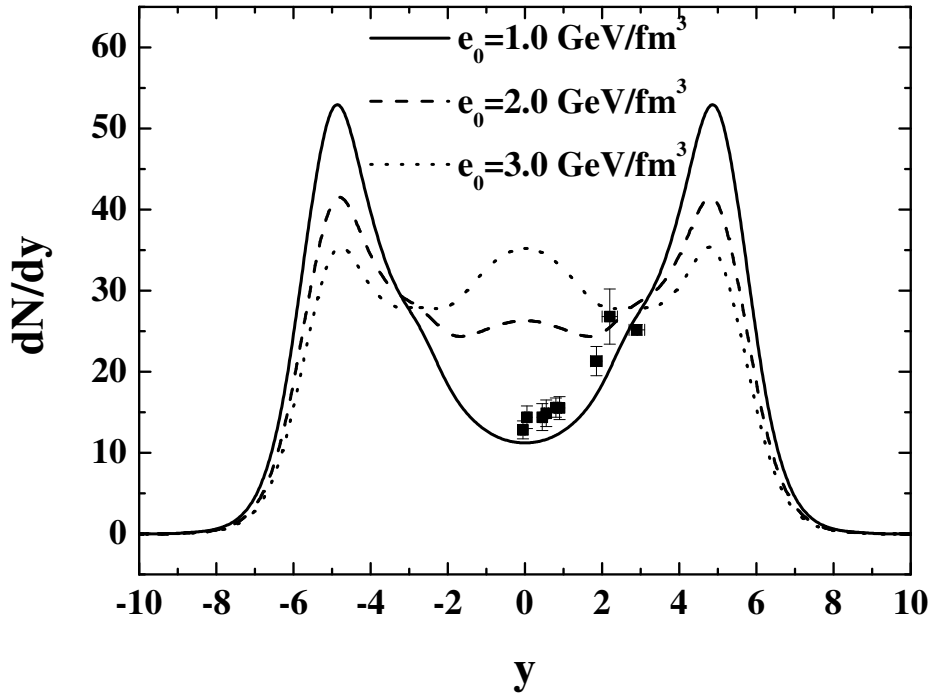} 
\vspace*{-0.5cm}
\caption{Net-baryon rapidity distribution in central Au+Au collisions ($N_{\rm part}$
=380) at $\sqrt{s_{NN}}=200$ GeV per $NN$-pair calculated for 3 different values of 
the parameter $\epsilon_0$ (indicated in the figure). Back reaction of the partonic 
plasma is not included in the calculation of the slab rapidities. Dots are 
experimental data of BRAHMS collaboration \cite{BRAHMS}.}
\end{figure}

The results shown in Fig.~3 were obtain without back reaction of the partonic
plasma on the baryon deceleration. As one can see, in this case we can get a 
perfect fit of the data by choosing $\epsilon_0\approx $ 1 GeV/fm$^3$, which, 
according to eq. (\ref{field}), corresponds to the actual field energy density 
of about 35 GeV/fm$^3$. Inclusion of the back reaction of plasma would require 
increasing of the initial chromofield energy density or/and the field decay time.   

On the basis of our model we can make predictions for the net-baryon dynamics 
for the future LHC experiments. We use eq. (\ref{field}) to extrapolate the 
initial field energy density from the RHIC ($\sqrt{s_0}$=200 GeV) to the LHC 
($\sqrt{s}$=5500 GeV) energy domain.  Then we get the mean net-baryon rapidity loss 
between 2.7 and 5.5 units for the calculation with and without the back reaction 
of partonic plasma, respectively. 


In conclusion, the collective deceleration of valence (net) baryons was studied 
assuming that strong longitudinal chromofields are formed at early stages 
of an ultrarelativistic heavy-ion collision.
We have solved classical equations of motion for baryonic slabs under
the action of a time-dependent chromoelectric field.A simpl;e hydrodynamical 
consideration shows that the energy density of produced plasma, due to the 
delayed production, is always significantly lower than the initial energy 
density of the chromofield. It has been demonstrated that 
the net-baryon rapidity loss $\langle\delta y\rangle\approx 2$ can be achieved with 
the initial energy density of the chromofield in the range of 40 to 70 GeV/fm$^3$.
The calculated net baryon rapidity distributions for central Au+Au collisions at RHIC
energies are in good agreement with BRAHMS data. The mean net-baryon rapidity loss for 
central Pb+Pb collisions at LHC energies is predicted between 2.7 and 5.5 units.

More detailed results, including the rapidity densities of net baryons and partonic 
plasma, calculated for different field decay patterns and centrality classes, will be 
presented in the forthcoming publication \cite{Mis2}. 

The authors thank C. Greiner, M. Gyulassy, L. McLerran, L.M. Satarov and H. St\"ocker 
for useful discussions. This work was supported in part by the DFG grant 436RUS 113/711/0-2 
(Germany), and grants RFFR-05-02-04013 and NS-8756.2006.2. (Russia).

\end{document}